\documentclass[pra,twocolumn,amsmath,amssymb,superscriptaddress,showpacs]{revtex4}

\usepackage{graphics}

\usepackage{epsfig}

\usepackage{bbm}
\usepackage[latin1]{inputenc} % Umlaute!

\newcommand{\be}{\begin{equation}}
\newcommand{\ee}{\end{equation}}
\newcommand{\eea}{\end{eqnarray}}
\newcommand{\bea}{\begin{eqnarray}}

\newcommand{\ex}[1]{\ensuremath{\left\langle{#1}\right\rangle}}
\newcommand{\exs}[1]{\ensuremath{\langle{#1}\rangle}}
\newcommand{\mean}[1]{\ensuremath{\langle{#1}\rangle}}

\newcommand{\eins}{\ensuremath{\openone}}
\newcommand{\qed}{\ensuremath{\hfill \Box}}
\newcommand{\WW}{\ensuremath{\mathcal{W}}}
\newcommand{\MM}{\ensuremath{\mathcal{M}}}
\newcommand{\PP}{\ensuremath{\mathcal{P}}}

\newcommand{\GEN}[1]{\ensuremath{\langle{#1}\rangle}}

\newcommand{\ketbra}[1]{\ensuremath{| #1 \rangle \langle #1 |}}

\newcommand{\ket}[1]{\ensuremath{|#1\rangle}}

\newcommand{\bra}[1]{\ensuremath{\langle#1|}}

\newcommand{\kommentar}[1]{}
\newcommand{\EQ}[1]{Eq.~(\ref{#1})}
\renewcommand{\vr}{\ensuremath{\varrho}}

\newcommand{\WGHZN}{\ensuremath{\mathcal{W}^{(GHZ_N)}}}
\newcommand{\WGHZTHREE}{\ensuremath{\mathcal{W}^{(GHZ_3)}}}
\newcommand{\WPGHZN}{\ensuremath{\widetilde{\mathcal{W}}^{(GHZ_N)}}}
\newcommand{\WGHZNPRIME}{\ensuremath{\widehat{\mathcal{W}}^{(GHZ_N)}}}
\newcommand{\WPGHZTHREE}{\ensuremath{\widetilde{\mathcal{W}}^{(GHZ_3)}}}
\newcommand{\WCN}{\ensuremath{\mathcal{W}^{(C_N)}}}
\newcommand{\WCNPRIME}{\ensuremath{\widehat{\mathcal{W}}^{(C_N)}}}
\newcommand{\WPCN}{\ensuremath{\widetilde{\mathcal{W}}^{(C_N)}}}
\newcommand{\WMERMIN}{\ensuremath{\mathcal{W}_{\rm{Mermin}}}}
\newcommand{\WTHREE}{\ensuremath{\mathcal{W}^{(W_3)}}}
\newcommand{\WPTHREE}{\ensuremath{\widetilde{\mathcal{W}}^{(W_3)}}}

\begin{document}
\title{Entanglement Detection in the Stabilizer Formalism}
\date{\today}
\begin{abstract}
We investigate how stabilizer theory can be used for constructing
sufficient conditions for entanglement. First, we show how
entanglement witnesses can be derived for a given state, provided
some stabilizing operators of the state are known. These witnesses
require only a small effort for an experimental implementation and
are robust against noise. Second, we demonstrate that also
nonlinear criteria based on uncertainty relations can be derived
from stabilizing operators. These criteria can sometimes improve
the witnesses by adding nonlinear correction terms. All our
criteria detect states close to Greenberger-Horne-Zeilinger
states, cluster and graph states. We show that similar ideas can
be used to derive entanglement conditions for states which do not
fit the stabilizer formalism, such as the three-qubit W state. We
also discuss
connections between the witnesses and
some Bell inequalities.
\end{abstract}

\author{G\'eza T\'oth}
\email{toth@alumni.nd.edu} \affiliation{Theoretical Division, Max
Planck Institute for Quantum Optics, Hans-Kopfermann-Stra{\ss}e 1,
D-85748 Garching, Germany.}

\author{Otfried G\"uhne}
\email{otfried.guehne@uibk.ac.at} \affiliation{Institut f\"ur
Quantenoptik und Quanteninformation, \"Osterreichische Akademie
der Wissenschaften,  A-6020 Innsbruck, Austria}

\pacs{03.67.Mn, 03.65.Ud, 03.67.-a}

\maketitle

% 03.65.Ud                  Entanglement in Quantum Mechanics
% 03.67.Mn                  Entanglement in Quantum Information
% 03.67.-a                  Quantum information

\section{Introduction}

Entanglement lies at the heart of quantum mechanics and plays also
a crucial role in quantum information theory. While the properties
of bipartite entanglement are still not fully understood, the
situation for multipartite entanglement is even more unclear,
since in the multipartite setting several inequivalent types of entanglement
occur. However, entangled states of many qubits are needed for
quantum information tasks such as measurement based quantum
computation \cite{RB03}, error correction \cite{G96,G97} or
quantum cryptography \cite{crypto}, to mention only a few. Thus it
is important both theoretically and experimentally to study
multipartite entanglement and to provide efficient methods to
verify that in a given experiment entanglement is really present.

In this paper we will apply the {\it stabilizer theory}
\cite{G96,G97} for entanglement detection. This theory already
plays a determining role in quantum information science. Its key
idea is describing the quantum state by its so-called {\it
stabilizing operators} rather than the state vector. This works as
follows: An observable $S_k$ is a stabilizing operator of an
$N$-qubit state $\ket{\psi}$ if the state $\ket{\psi}$ is an
eigenstate of $S_k$ with eigenvalue $1$
\begin{equation}
S_k \ket{\psi}=\ket{\psi}. \label{stabil}
\end{equation}
Many highly entangled $N$-qubit states can be uniquely defined by
$N$ stabilizing operators which are locally measurable, i.e., they
are products of Pauli matrices.

The main result of the present paper can be formulated as follows:
If one has a given state $\ket{\psi}$ and has identified some of
its stabilizing operators, then it is easy to derive entanglement
conditions which detect states in the proximity of $\ket{\psi}.$
So looking for stabilizing operators should be the first step in
order to detect entanglement.
All the conditions presented
are easy to implement
in experiments and are robust against noise. We mainly consider
criteria detecting entanglement close to
Greenberger-Horne-Zeilinger (GHZ) \cite{GH90}, cluster \cite{BR03}
and graph states \cite{HE04}.
We use different types of entanglement conditions, they may be
linear criteria, such as entanglement witnesses, or nonlinear
criteria, based on uncertainty relations. In this way we complete
our results of Ref.~\cite{TG04}, where witnesses for the detection
of multipartite entanglement in the vicinity of GHZ and cluster
states were derived. Note that the stabilizer formalism also
allows to derive Bell inequalities, this has been recently
investigated in Ref.~\cite{SA04}. We do not aim to derive Bell
inequalities here, since they are used to rule out local hidden
variable models, a notion independent of quantum physics. However,
some of our constructions exhibit close connections to Bell
inequalities, and this will also be discussed.

Our paper is organized as follows. Since GHZ states are the most
studied stabilizer states, we use mainly them to explain our
ideas, the generalization to other stabilizer states is then
usually straightforward. So we start in Section II by recalling
the basic facts about the stabilizing operators of GHZ states.
Then we present a method for obtaining a
family of entanglement witnesses for detecting entanglement close to GHZ
states. First we present witnesses
detecting any (i.e., even partial or biseparable) entanglement.
Then we present witnesses which detect only genuine multi-qubit
entanglement. We discuss some interesting connections to Bell
inequalities. In Section III we present witnesses for cluster and
graph states. We also consider detecting entanglement close to
given mixed states. Finally, we present entanglement witnesses for
a W state. It is of interest since the W state does not fit the
stabilizer framework. We show that our ideas can still be
generalized for this case. In Section IV we present nonlinear
entanglement conditions in the form of variance based uncertainty
relations. It turns out that they can often improve the witnesses
by adding nonlinear terms. In Section V we present entanglement
conditions which are based on entropic uncertainty relations. Finally, in
the Appendix we collect some basic facts about the stabilizer
formalism and present some technical calculations in detail.

\section{GHZ states as examples of stabilizer states}

Let us start by introducing the stabilizer formalism using the
example of GHZ states. An $N$-qubit GHZ state is given by
\begin{equation}
\ket{GHZ_N}=\frac{1}{\sqrt{2}}
(\ket{0}^{\otimes N}+\ket{1}^{\otimes N}).
\end{equation}
Besides this explicit definition one may define the GHZ state also
in the following way: Let us look at the observables
\begin{eqnarray}
S_1^{(GHZ_N)}&:=& \prod_{k=1}^N X^{(k)},
\nonumber\\
S_k^{(GHZ_N)}&:=&Z^{(k-1)} Z^{(k)} \mbox{ for } k=2,3,...,N,
\label{eigenGHZ}
\end{eqnarray}
where $ X^{(k)}, Y^{(k)}$, and $ Z^{(k)}$ denote the Pauli
matrices acting on the $k$-th qubit. Now we can {\it define} the GHZ
state as the state $\ket{GHZ_N}$ which fulfills
\begin{equation}
S_k^{(GHZ_N)}\ket{GHZ_N}=\ket{GHZ_N} \label{eigenGHZ2}
\end{equation}
for $k=1,2,...,N.$ One can straightforwardly calculate that these
definitions are equivalent and the GHZ state is uniquely defined
by the Eqs.~(\ref{eigenGHZ2}). From a physical point of view the
definition via Eqs.~(\ref{eigenGHZ2}) stresses that the GHZ state
is uniquely determined by the fact that it exhibits perfect
correlations for the observables $S_k^{(GHZ_N)}.$

Note that $\ket{GHZ_N}$ is stabilized not only by $S_k^{(GHZ_N)}$,
but also by their products. These operators, all having perfect
correlations for a GHZ state, form a group called {\it stabilizer}
\cite{G96}. This $2^N$-element group of operators will be denoted
by $\mathcal{S}^{(GHZ_N)}$. $S_k^{(GHZ_N)}$ are the generators of
this group which we will denote as
$\mathcal{S}^{(GHZ_N)}=\GEN{S_1^{(GHZ_N)},S_2^{(GHZ_N)},...,S_N^{(GHZ_N)}}.$
For more details on the stabilizer please see Appendix A.

\subsection{Witnesses for stabilizer states}

In order to show that a given state contains some entanglement, we
have to exclude the possibility that the state is fully separable,
i.e., it can be written as
\begin{equation}
\vr=\sum_i p_i \vr_i^{(1)} \otimes \vr_i^{(2)} \otimes ... \otimes
\vr_i^{(N)} \label{fullsep}
\end{equation}
with $p_i \geq 0,\sum_i p_i =1.$

Before presenting entanglement witness operators,
let us shortly recall their definition.
An entanglement witness $\WW$ is an observable which has a positive or zero
expectation value for all separable states, and a negative one on some
entangled states \cite{HH90}:
\begin{equation}
Tr(\WW\varrho)
\left\{
\begin{array}{l}
\geq 0 \,\, \textrm{ for all separable states }\varrho_s.
\\
<0 \,\, \textrm{ for some entangled states }\varrho_e.
\end{array}
\right.
\end{equation}
Thus a negative expectation value in an experiment signals the presence
of entanglement.

In this paper we will construct a family of entanglement witnesses
using the elements of the stabilizer. We will call these {\it
stabilizer witnesses}.

\subsection{Ruling out full separability}

In the following stabilizer witnesses will be used to
detect entanglement close to GHZ states.
We will construct witnesses of the form
\begin{equation}
\WW :=c_0\eins-\widetilde{S}_k^{(GHZ_N)}-\widetilde{S}_l^{(GHZ_N)}, \label{wstabil2}
\end{equation}
where $\widetilde{S}_{k/l}^{(GHZ_N)}$ are elements of the stabilizer group,
\begin{equation}
c_0:=\max_{{\vr}\in \mathcal{P}} \big[
\exs{\widetilde{S}_k^{(GHZ_N)}+\widetilde{S}_l^{(GHZ_N)}}_\vr \big].
\end{equation}
and $\mathcal{P}$ denotes the set of product states. Since
the set of separable states is convex, $c_0$ is also the maximum
for mixed separable states of the form Eq.~(\ref{fullsep}).

Clearly, if we want to detect entangled states with $\WW$, we have
to choose $\widetilde{S}_k^{(GHZ_N)}$ and $\widetilde{S}_l^{(GHZ_N)}$ such
that the maximum of
$\exs{\widetilde{S}_k^{(GHZ_N)}+\widetilde{S}_l^{(GHZ_N)}}$ for entangled
quantum states is larger than the maximum for separable states.
Whether this condition holds depends on the question whether the
$\widetilde{S}_k^{(GHZ_N)}$ and $\widetilde{S}_l^{(GHZ_N)}$ commute
locally:
\\
{\it
{\bf Definition 1.} Two correlation operators of the
form
\begin{eqnarray}
K&=&K^{(1)}\otimes K^{(2)}\otimes ... \otimes K^{(N)},\nonumber\\
L&=&L^{(1)}\otimes L^{(2)}\otimes ... \otimes L^{(N)}
\label{loccom}
\end{eqnarray}
commute {\it locally} if
}
\begin{equation}
\text{for every } n\in\{1,2,...,N\} :
K^{(n)}L^{(n)}=L^{(n)}K^{(n)}.
\end{equation}
Using Definition 1, we can make the following
statement:
\\
{\it
{\bf Observation 1.} Two multi-qubit correlation operators, $K$ and
$L$, commute locally iff there is a pure product
state among their common eigenstates. 
}
\\
{\it Proof.} $K$ and $L$ commute locally iff for all $n\in\{1,2,...,N\}$
there are two vectors, $\ket{\phi_n}$ and $\ket{\phi_n^\perp}$, which are common eigenstates of $K^{(n)}$ and $L^{(n)}$.
Thus $\ket{\psi}=\ket{\phi_1}\otimes \ket{\phi_2}\otimes ...
\otimes \ket{\phi_N}$ is a common eigenstate of $K$ and $L.$
$\qed$
Hence it follows
that if $\widetilde{S}_k^{(GHZ_N)}$ and $\widetilde{S}_l^{(GHZ_N)}$ commute locally then the maximum of
$\exs{\widetilde{S}_k^{(GHZ_N)}+\widetilde{S}_l^{(GHZ_N)}}$ for separable and entangled
states coincide.

After these considerations, we construct our witness from two
locally non-commuting stabilizing operators:
\\
{\it
{\bf Theorem 1.} A witness detecting entanglement
around an $N$-qubit GHZ state is
}
\begin{equation}
\WGHZN_m := \eins-S_1^{(GHZ_N)}- S_m^{(GHZ_N)},\label{fsepGHZ}
\end{equation}
{\it
where $m=2,3,...,N$.
}
\\
{\it Proof.}
The proof is based on the Cauchy-Schwarz inequality.
Using this and $\mean{X^{(i)}}^2+\mean{Z^{(i)}}^2 \leq
1$, for pure product states we obtain
\begin{align}
& \!\!\!\!\!\!\exs{S_1^{(GHZ_N)}} + \exs{S_m^{(GHZ_N)}}=
\nonumber \\
&= \exs{X^{(1)}}\exs{X^{(2)}}...\exs{X^{(N)}} +
\exs{Z^{(m-1)}}\exs{Z^{(m)}}
\nonumber\\
&\le |\exs{X^{(m-1)}}|\cdot|\exs{X^{(m)}}|+ |\exs{Z^{(m-1)}}|
\cdot|\exs{Z^{(m)}}|
\nonumber\\
&\le \sqrt{\exs{X^{(m-1)}}^2+\exs{Z^{(m-1)}}^2 }
\sqrt{\exs{X^{(m)}}^2+\exs{Z^{(m)}}^2 }
\nonumber\\
&\le 1. \label{cschwarz}
\end{align}
It is easy to see that the bound is also valid for mixed separable states.
This proof can straightforwardly be generalized for arbitrary two
locally non-commuting elements of the stabilizer. $\qed$

Witnesses can be constructed with more than two elements of the
stabilizer as
\begin{equation}
\WGHZNPRIME_m:=\eins - S_1^{(GHZ_N)}- S_m^{(GHZ_N)} -
S_1^{(GHZ_N)}S_{m}^{(GHZ_N)} \label{fsepGHZprime}
\end{equation}
for $m =2,3,...,N$ rule out full separability. This can be proved
in a similar calculation as in Theorem 1, using the
fact that $\mean{X^{(i)}}^2+\mean{Y^{(i)}}^2+\mean{Z^{(i)}}^2 \leq
1.$ Later, in Sec. IV we will see how these conditions can be
improved by adding refinement terms which are quadratic in the
expectation values.

\subsection{Criteria for witnesses}

Having derived the first entanglement witnesses, it is now
time to ask, whether they are really useful witnesses.
In fact, there are, for a given state, always infinitely
many witnesses allowing the detection of this state.
However, some of them are more useful than others. Two
criteria for the usefulness of witnesses are of interest:
Firstly, it is important to characterize how much a witness
tolerates noise. Secondly, it is crucial to know, how much
experimental effort is needed to measure the witness.

Witnesses are usually designed to detect entangled states close to
a given pure state $\ket{\Psi}$. From a practical point of view it
is very important to know, how large neighborhood of
$\ket{\Psi}$ is detected as entangled. This can be
characterized in the following way:
\\
{\it
{\bf Definition 2.} Let the density matrix of the state obtained after
mixing with white noise be given by
}
\begin{equation}
\varrho(p_{\rm noise}) := p_{\rm noise} \frac{\eins}{2^N} + (1-p_{\rm noise})
\ketbra{\Psi}.
\label{mixed_with_noise}
\end{equation}
{\it
Here $p_{\rm noise}$ determines the ratio of white noise in the
mixture. Then, the {\rm robustness to noise} for a witness $\WW$ is
determined by the maximal noise ratio for which it still detects
$\varrho(p_{\rm noise})$ as entangled.
}

It is easy to see that
witness $\WW$ detects $\vr(p_{\rm noise})$ as entangled if $p_{\rm
noise}<p_{\rm limit}$ where
\begin{equation}
p_{\rm limit}:=\frac{-\langle GHZ_N|\WW|GHZ_N\rangle}
{2^{-N}Tr(\WW)-\langle GHZ_N|\WW|GHZ_N\rangle}. \label{plimit2}
\end{equation}
To give an example, for witnesses of the form \EQ{wstabil2} we
have $p_{\rm limit}=(\gamma-1)/\gamma$ where
\begin{equation}
\gamma:=\frac{\langle
GHZ_N|S_k^{(GHZ_N)}+S_l^{(GHZ_N)}|GHZ_N\rangle} {\max_{{\vr}\in
\mathcal{P}} \big[ \exs{S_k^{(GHZ_N)}+S_l^{(GHZ_N)}}_\vr\big]}.
\end{equation}
Clearly, the maximum of the numerator is $2$, while the minimum of
the denominator is $1$. Thus the maximum noise tolerance achievable
by a witness of the form \EQ{wstabil2} is given by $p_{\rm
limit}=1/2.$ Witnesses $\WGHZN_{m}$ from Eq.~(\ref{fsepGHZ}) have
exactly this noise tolerance thus they are optimal among
stabilizer witnesses with two correlation terms. Witnesses
$\WPGHZN_{m}$ from Eq.~(\ref{fsepGHZprime}) detect  $\vr(p_{\rm
noise})$ as entangled if $p_{\rm noise}<2/3,$ thus, they are more
robust against noise. It can be proved that
Eq.~(\ref{fsepGHZprime}) is optimal among stabilizer witnesses
with three correlations terms from the point of view of noise
tolerance.

The experimental effort needed for measuring a witness can be
characterized by the number of {\it local measurement settings}
needed for its implementation \cite{exdet}:
\\
{\it
{\bf Definition 3.}
The local decomposition of a witness is defined as follows:
Any witness can be decomposed  into a sum
$\WW=\sum_i \MM_i$ where each of the terms $\MM_i$
can be measured by a local measurement setting.
One local setting
$ \mathcal{L} = \{O^{(k)}\}_{k=1}^N$ consists of performing simultaneously
the von Neumann measurements $O^{(k)}$ on the corresponding parties.
By repeating the measurements many times one can determine the
probabilities of the $2^N$ different outcomes. Given these
probabilities it is possible to compute all two-point correlations
$\exs{O^{(k)} O^{(l)}}$, three-point correlations
$\exs{O^{(k)}O^{(l)}O^{(m)}},$ etc. }

It is easy to see that measuring only one setting is
not enough for entanglement detection.
For measuring all the witnesses $\WGHZN_{m}$ given
Eq.~(\ref{fsepGHZ}) for $m=2,3,...,N$, two measurement settings
are required for an implementation, namely
$\{X^{(1)},X^{(2)},...,X^{(N)}\}$ and
$\{Z^{(1)},Z^{(2)},...,Z^{(N)}\}.$ The witnesses $\WGHZNPRIME_{m}$
from Eq.~(\ref{fsepGHZprime}) require a measurement of three
settings: $\{X^{(1)},X^{(2)},...,X^{(N)}\},$
$\{Y^{(1)},Y^{(2)},...,Y^{(N)}\}$ and
$\{Z^{(1)},Z^{(2)},...,Z^{(N)}\}.$

\subsection{Detecting genuine multipartite entanglement}

Up to now we considered witnesses which detect any (even partial)
entanglement. However, for multipartite systems there are several
classes of entanglement. The most interesting class of entangled
states are the {\it genuine multipartite entangled} states. These
are defined as follows. Let us assume that a pure state
$\ket{\psi}$ on an $N$ qubit system is given. If we can find a
partition of the $N$ qubits into two groups $\mathcal{A}$ and
$\mathcal{B}$ such that $\ket{\psi}$ is a product state with
respect to this partition,
\begin{equation}
\ket{\psi} = \ket{\phi}_\mathcal{A}
\ket{\chi}_\mathcal{B}
\end{equation}
then we call the state $\ket{\psi}$ {\it biseparable} (with
respect to the given partition). Note that the states
$\ket{\phi}_\mathcal{A}$ and $\ket{\chi}_\mathcal{B}$ may be
entangled, thus the state $\ket{\psi}$ is not necessarily fully
separable. According to the usual definition a mixed state is
called biseparable iff it can be written as a convex combination
of pure biseparable states
\begin{equation}
\vr=\sum_i p_i \ketbra{\psi_i}
\end{equation}
where the $\ket{\psi_i}$ may be biseparable with respect to
different partitions. If a state is not biseparable then 
it is called {\it genuine
multipartite entangled.} In experiments dealing with the
generation of entanglement in multi-qubit systems it is necessary
to generate and verify genuine multipartite entanglement, since
the simple statement ``The state is entangled'' would still allow
that only two of the qubits are entangled while the rest is in a
product state.

Witnesses for the detection of genuine multipartite entanglement
have already been used experimentally for entanglement detection
close to three-qubit W states and four-qubit singlet states
\cite{BE03}. They used the projector on the state to be detected
as an observable to detect entanglement.
For GHZ states, the projector-based witness reads
\begin{equation}
\WPGHZN:=\frac{1}{2}\eins -\ket{GHZ_N}\bra{GHZ_N},
\label{GHZWITNESS}
\end{equation}
which has also been used in an experiment \cite{SK00}.
This witness detects genuine $N$-qubit entanglement around an
$N$-qubit GHZ state. The constant $1/2$ corresponds to the maximal
squared overlap  between the GHZ state and the pure biseparable
states, this can be calculated using the methods presented in
Ref.~\cite{BE03}. 

Witness Eq.~(\ref{GHZWITNESS}) can be interpreted in the following
way. $\exs{\WPGHZN}$ is $-1/2$ only for the GHZ state state. For
any other state it is larger. In general, the more negative
$\exs{\WPGHZN}_{\ket{\Psi}}$ is, the closer state $\ket{\Psi}$ is
to the GHZ state. It is known that in the proximity of the GHZ
state there are only states with genuine $N$-qubit entanglement,
so the constant in Eq.~(\ref{GHZWITNESS}) is chosen such that if
$\exs{\WPGHZN}<0$ then the state is in this neighborhood and is
detected as entangled.

Now we will show that the projector-based witness
Eq.~(\ref{GHZWITNESS}) can be constructed as a sum of elements of
the stabilizer. First, note that the generators of the stabilizer
$S_k^{(GHZ_N)}$ can be used to define a very convenient basis for
calculations, namely their common eigenvectors. $\ket{GHZ_N}$ is
one of them giving a $+1$ eigenvalue for all $S_k^{(GHZ_N)}$'s.
Allowing both $+1$ and a $-1$ eigenvalues in Eq.~(\ref{stabil}),
$2^N$ $N$-qubit states can be defined which are orthogonal to each
other and form a complete basis. These are all GHZ states, but in
different bases, i.e., they are of the form $(\ket{x^{(1)} ...
x^{(N)}}\pm \ket{\overline{x}^{(1)} ...
\overline{x}^{(N)}})/\sqrt{2} $ with $x^{(l)} \in \{0,1\}$ and
$\overline{x}^{(l)} = 1 - x^{(l)}.$ We will refer to this basis as
the {\it GHZ state basis}.
This basis will turn
out to be extremely useful since all operators of the stabilizer
are diagonal in this basis.

Now, the projector onto a GHZ state
can be written with the stabilizing operators \cite{STABIL}
\begin{equation}
\ket{GHZ_N}\bra{GHZ_N} = \prod_{k=1}^N
\frac{S_k^{(GHZ_N)}+\eins}{2}. \label{PGHZ}
\end{equation}
This can be directly seen in the GHZ basis. From Eq.~(\ref{PGHZ})
it follows that the projector-based witness can also be decomposed
into the sum of stabilizing terms, i.e., $\WPGHZN$ is a stabilizer
witness. This decomposition can be used to measure the witness
locally. However, the number of settings needed seems to increase
exponentially with the number of qubits \cite{GH03}. Thus we have
to find other stabilizer witnesses for which measuring them is
more feasible.

For detecting biseparable entanglement it was enough to use two
stabilizing operators for the witness. For the detection of
$N$-qubit entanglement we have to make measurements on {\it all}
qubits and have to measure a full set of generators. This is because we
need that the expectation value of the witness is
minimal only for the GHZ state.
If the witness does not contain a full set of generators then
there are at least two of the elements of the GHZ basis
giving the minimum. There is, however, always a superposition of
two basis vectors of the GHZ basis which is biseparable.
This biseparable state would also give a minimum for our witness.
Thus this witness could not be used for detecting genuine $N$-qubit
entanglement.

The main idea of detecting genuine multi-qubit entanglement with
the stabilizing operators is the following:
\\
{\it
{\bf Observation 2.} Let us consider some of the stabilizing operators. If this
set of operators contains a complete set of generators and for a
given state the expectation values of these correlation operators
are close enough to the values for a GHZ state, then this state
must be close to a GHZ state and is therefore multi-qubit entangled.
}

Now we can derive the first entanglement witness.
\\
{\it
{\bf Theorem 2.}
The witness
}
\begin{equation}
\WW:= (N-1) \eins -\sum_{k=1}^{N} S_k^{(GHZ_N)}
\label{GHZN}
\end{equation}
{\it
detects only states with genuine $N$-qubit entanglement.
}
\\
{\it Proof.} In Eq.~(\ref{GHZN}) the constant term, $c_0=N-1$,
was chosen such that the observable
\begin{equation}
X_\alpha:=\WW-\alpha\WPGHZN\ge 0.
\label{posdef}
\end{equation}
for some $\alpha>0$ is positive semidefinite. Then we have for any
state $\varrho$ the inequality $\alpha Tr(\varrho \WPGHZN) \le
Tr(\varrho{\WW})$ which implies that $Tr(\varrho{\WW})< 0$ can
only happen if $Tr(\varrho \WPGHZN < 0.$ Thus all states detected
with $\WW$ are also detected by $\WPGHZN$ and ${\WW}$ is a
multi-qubit witness. Clearly, we would like to have $c_0$ as small
as possible, since the smaller $c_0$ is, the more entangled states
$\WW$ detects. Simple calculation leads to $c_0=N-1$. One can
check that with this choice $\WW-2\WPGHZN\ge 0$. $\qed$

\begin{figure}
%\narrowtext
\centerline{\epsfxsize=2.0in\epsffile{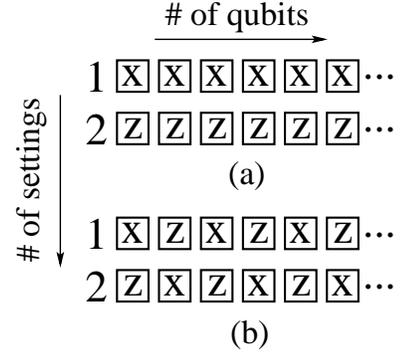}
} \caption{
  (a) Measurement settings needed for the approach presented
  in this paper for detecting entangled
  states close to GHZ states and (b) cluster states.
  For each qubit the measured spin component is indicated.}
\label{fig_settings}
\end{figure}

Both witnesses  $\WPGHZN$ and ${\WW}$ detect entangled states
close to GHZ states. The main advantage of ${\WW}$ in comparison
with $\WPGHZN$ lies in the fact that for implementing it {\it only
two} measurement settings are needed, namely the ones shown in
Fig.~\ref{fig_settings}(a). The first setting,
$\{X^{(1)},X^{(2)},...,X^{(N)}\},$ is needed to measure
$\exs{S_1^{(GHZ_N)}}$, the second one,
$\{Z^{(1)},Z^{(2)},...,Z^{(N)}\}$, is to measure the expectation
values for the other generators. The other characteristic to check
is the noise tolerance of the witness. The witness ${\WW}$ detects
states mixed with noise of the form \EQ{mixed_with_noise} if
$p_{\rm noise} < 1/N$. Thus the noise tolerance decreases as the
number of qubits increases.

However, using a similar construction it is also possible to
derive a witness which is robust against noise even for many
qubits and still requires only two measurement
settings:
\\
{\it
{\bf Theorem 3.} The following entanglement witness detects genuine
$N$-qubit entanglement for states close to an $N$-qubit GHZ state
}
\begin{equation}
\WGHZN := 3\eins - 2\bigg[\frac{S_1^{(GHZ_N)}+\eins}{2}
+\prod_{k=2}^N\frac{S_k^{(GHZ_N)}+\eins}{2} \bigg].
\label{GHZNbetter}
\end{equation}
{\it
This witness has the best noise tolerance among stabilizer
witnesses which need only two measurement settings and have the
property $\WGHZN-\alpha\WPGHZN\ge 0$ for some $\alpha>0.$
}
\\
{\it Proof.}
To prove the first statement, one can  show by direct calculation that
$\WGHZN-2\WPGHZN\ge 0$.
Thus $\WGHZN$ is a multi-qubit witness.
For the proof of optimality please see Appendix B.
$\qed$

To give an example, for the simple case of three qubits the
witness is
\begin{eqnarray}
\WGHZTHREE&:=& \frac{3}{2}\eins
-X^{(1)}X^{(2)} X^{(3)}\nonumber\\
&-&\frac{1}{2}\big[Z^{(1)}Z^{(2)} +Z^{(2)}Z^{(3)}
+Z^{(1)}Z^{(3)}\big]. \nonumber\\
\end{eqnarray}
Three-qubit genuine multi-qubit entangled states can belong to the
so-called W-class or to the GHZ-class \cite{AB01}.
Knowing that $\WPGHZTHREE+\eins/4$ detects GHZ-class entanglement
\cite{AB01}, we obtain that $\WGHZTHREE+\eins/2$ detects also only
GHZ-class entanglement.

The witness $\WGHZN$ is quite robust against noise. It detects states
mixed with white noise as true multipartite entangled for
\begin{equation}
p_{\rm noise}<\frac{1}{3-2^{(2-N)}}.
\end{equation}
Thus it tolerates at least $33\%$
noise, independently from the number of qubits. Again, only two
measurement settings are necessary for an implementation [see
Fig.~\ref{fig_settings}(a)].

The expression for the witness $\WGHZN$ can be simplified using
the fact that
\begin{equation}
\prod_{k=2}^N \frac{S_k^{(GHZ_N)}+\eins}{2} =
\ketbra{00...0} + \ketbra{11...1},
\end{equation}
where $\ketbra{11...1}$ is
the projector on the state with all spins down and
$\ketbra{00...0}$ is the projector on the state with all spins up.
Using this one obtains
\begin{eqnarray}
\WGHZN &=& 2\eins -
X^{(1)}X^{(2)}X^{(3)} \cdot \cdot \cdot X^{(N)}
\nonumber \\
&&
- 2\ketbra{00...0} -  2\ketbra{11...1}.
\end{eqnarray}

\subsection{Bell inequalities for GHZ states}

As a sidestep we will discuss now a very surprising feature of the
stabilizer witnesses, namely that they are closely related to
Mermin-type Bell inequalities \cite{B64,M90}. As we will see, this
relationship sheds new light on the question whether and when Bell
inequalities \cite{SU01,GB98,C02} can detect genuine multipartite
entanglement.

First note that witnesses different from the previous ones
can be obtained by
including further terms of the stabilizer and using
more than two measurement settings. For instance,
following the lines of the previous section it is
easy to see that the observable
\begin{eqnarray}
\WMERMIN&:=& 2\eins - S_1^{(GHZ_3)} [\eins+
S_2^{(GHZ_3)}][\eins+S_3^{(GHZ_3)}]\nonumber\\
&=& 2\eins + Y^{(1)}Y^{(2)}X^{(3)} +
X^{(1)}Y^{(2)}Y^{(3)}\nonumber\\
&+& Y^{(1)}X^{(2)}Y^{(3)} - X^{(1)}X^{(2)}X^{(3)}
\end{eqnarray}
detects genuine three-party entanglement around a GHZ state. It
detects a GHZ state mixed with white noise if $p_{\rm noise}<1/2$.
The witness $\WMERMIN$  is equivalent to
Mermin's inequality \cite{M90}
\begin{eqnarray}
\mean{X^{(1)}X^{(2)}X^{(3)}} -
\mean{X^{(1)}Y^{(2)}Y^{(3)}} &-&
\nonumber \\
\mean{Y^{(1)}X^{(2)}Y^{(3)}} - \mean{Y^{(1)}Y^{(2)}X^{(3)} } &\leq
&2.
\label{Sigma1}
\end{eqnarray}
In Ref.~\cite{PB00} the condition given in Eq.~(\ref{Sigma1}) was
used for detecting entanglement in a $3$-qubit photonic system and
a measurement result equivalent to $\exs{\WMERMIN} \approx
-0.83\pm0.09$ was obtained, thus the created state was genuine
three-qubit entangled.

For $N>3$ the Bell operator in Mermin's inequality contains also
only stabilizing terms:
\\
{\it
{\bf Theorem 4.} For the Bell operator of the Mermin's inequality
\cite{M90B,Nagata}
\begin{eqnarray}
M_N&:=&\frac{1}{2^{N-1}} \bigg[ X^{(1)}X^{(2)}X^{(3)}X^{(4)}
\cdot \cdot \cdot X^{(N-1)}X^{(N)}\nonumber\\
&-& Y^{(1)}Y^{(2)}X^{(3)}X^{(4)} \cdot \cdot \cdot X^{(N-1)}X^{(N)}\nonumber\\
&+&Y^{(1)}Y^{(2)}Y^{(3)}Y^{(4)}\cdot \cdot \cdot X^{(N-1)}X^{(N)}
- ...\bigg].\nonumber\\\label{MMM}
\end{eqnarray}
the operator expectation value for biseparable states is bounded
by
\begin{eqnarray}
\exs{M_N}\le \frac{1}{2},
\end{eqnarray}
while the quantum maximum is $1.$ Note that a term in
Eq.~(\ref{MMM}) represents the sum of all its possible
permutations.
}
\\{\it Proof.} $M_N$ can
alternatively written as
\begin{eqnarray}
M_N&=&S_1^{(GHZ_N)}\prod_{k=2}^{N} \frac{S_k^{(GHZ_N)}+\eins}{2}\nonumber\\
&=&\ket{00...0}\bra{11...1}+\ket{11...1}\bra{00...0}.
\end{eqnarray}
The maximum for $\exs{M_N}$ for biseparable states can be obtained knowing
\begin{equation}
\ketbra{GHZ_N}-M_N=\ketbra{GHZ_N^-}\ge 0,
\end{equation}
where $\ket{GHZ_N^-}=(\ket{0000...}-\ket{1111...})/\sqrt{2}.$
Hence for biseparable states $\varrho$
\begin{equation}
\exs{M_N}_\varrho \le \ex{\ketbra{GHZ_N}}_\varrho\le \frac{1}{2}.
\end{equation}
For fully separable states the bound is lower and is given in
Ref.~\cite{R05}. $\qed$

\section{Witnesses for cluster, graph and W states}

\subsection{Witnesses for cluster states}

Let us turn now to cluster states. These are  a family of
multi-qubit states which have attracted increasing attention in
the last years. A cluster state $\ket{C_N}$ is defined to be the
state fulfilling the equations $\ket{C_N}= S_k^{(C_N)} \ket{C_N}$
with the following stabilizing operators
\begin{eqnarray}
S_1^{(C_N)}&:=&X^{(1)}Z^{(2)},
\nonumber\\
S_k^{(C_N)}&:=&Z^{(k-1)} X^{(k)} Z^{(k+1)}; k=2,3,...,N-1,
\nonumber\\
S_N^{(C_N)}&:=&Z^{(N-1)} X^{(N)}. \label{eigenC}
\end{eqnarray}

Witnesses similar to \EQ{fsepGHZ} which rule out full separability
can be constructed with two locally non-commuting operators as
\cite{T03}
\begin{equation}
\WCN_k:= \eins-S_k^{(C_N)}-S_{k+1}^{(C_N)} \mbox{ for } k
=1,2,...,N-1.\label{fsepC}
\end{equation}
This witness detects biseparable
states close to an $N$-qubit cluster state. The proof is
essentially the same as the one for \EQ{fsepGHZ}.

Note that $\WCN_k$ involves only two generators which act on at
most four-qubits. This witness detects whether the reduced density
matrix of the qubit quadruplet corresponding to qubits
$(k-1),(k),(k+1)$ and $(k+2)$ is entangled. The state of the rest
of the qubits does not influence $\exs{\WCN_k}.$ The witness
$\WCN_k$ tolerates noise if $p_{\rm noise}<1/2$.

The following witnesses have a better noise tolerance
\begin{eqnarray}
\WCNPRIME_k&:=& \eins-S_k^{(C_N)}-S_{k+1}^{(C_N)}-
S_k^{(C_N)}S_{k+1}^{(C_N)}
\nonumber\\
&& \mbox{ for } k=1,2,...,N-1. \label{fsepC2}
\end{eqnarray}
This witness still involves only the qubits of a quadruplet and
tolerates noise if $p_{\rm noise}<2/3$.

Using witnesses $\WCNPRIME_k$, one can construct a ``composite''
entanglement witness for which the noise tolerance increases with
the number of qubits:
\\
{\it
{\bf Theorem 5.} The following
entanglement witness detects entangled states close to a cluster
state
}
\begin{equation}
\WW^{(C_N)}_{comp}:=\prod_{k=0}^{K-1}
\eins-S_{4k+1}^{(C_N)}-S_{4k+2}^{(C_N)}-
S_{4k+1}^{(C_N)}S_{4k+2}^{(C_N)}, \label{WWCNcomposite}
\end{equation}
{\it
where $K:=\lfloor (N+2)/4 \rfloor$ and
$\lfloor x\rfloor$ denotes the integer part of $x$. The witness
Eq.~(\ref{WWCNcomposite}) tolerates noise if \be p_{\rm
noise}<\frac{2^{K}}{2^{K}+1}.\ee
}
 {\it Proof.} The noise tolerance
comes from direct calculation along the lines of \EQ{plimit2}.
Note that all the terms in the product in
Eq.~(\ref{WWCNcomposite}) act on disjoint sets of qubits. This is
the reason that such a composite witness can be constructed.
$\qed$

One can also construct witnesses for the detection of genuine
multipartite entanglement close to cluster states, similar as for
the case of GHZ states:
\\
{\it
{\bf Theorem 6.} The witnesses
}
\begin{eqnarray}
\WPCN&:=&\frac{\eins}{2}-\ketbra{C_N},\label{PCN}\\
\WCN &:=& 3\eins - 2\bigg[\prod_{\text{odd
k}}\frac{S_k^{(C_N)}+\eins}{2}+
\prod_{\text{even k}}\frac{S_k^{(C_N)}+\eins}{2}\bigg].\nonumber\\
\label{CN}
\end{eqnarray}
{\it
detect genuine $N$-party entanglement close to a cluster state.
$\WCN$  is optimal from the point of view of noise tolerance among
the stabilizer witnesses which need only two measurement settings
and have the
property $\WCN-\alpha\WPCN\ge 0$ for some $\alpha>0.$
}
\\
{\it Proof.}
First we have to prove that Eq.~(\ref{PCN}) is a multi-qubit witness.
For that we have to use that from a cluster state one can
generate a singlet between  arbitrary qubits by local operations
\cite{BR03}. Using the results of Ref. \cite{N99} this implies
that the maximal Schmidt coefficient of a cluster state when making
a Schmidt decomposition with respect to an arbitrary bipartite
split does not exceed the maximal Schmidt coefficient of the
singlet, which equals $1/\sqrt{2}.$ 
Then, from the methods of
Ref.~\cite{BE03} it follows that $\WPCN$ is
a witness for multi-qubit entanglement. After that we have to prove that
Eq.~(\ref{CN}) is a multi-qubit witness.
This can be proved similarly as it has been done for Theorem 3
using that $\WCN-2\WPCN\ge 0.$
Concerning the optimality, see
Appendix B. $\qed$

The stabilizing operators  in $\WCN$ are again divided into two
groups corresponding to the two settings
$\{X^{(1)},Z^{(2)},X^{(3)},Z^{(4)},...\}$ and
$\{Z^{(1)},X^{(2)},Z^{(3)},X^{(4)},...\}$ as shown in
Fig.~\ref{fig_settings}(b). Witness $\WW^{(C_N)}$ tolerates
mixing with noise if
\begin{eqnarray}
p_{\rm noise}<\bigg\{
\begin{array}{ll}
(4-4/2^{\frac{N}{2}})^{-1} & \mbox{ for even }N \\
{[}4-2(1/2^{\frac{N+1}{2}}+1/2^{\frac{N-1}{2}})]^{-1} & \mbox{ for
odd } N
\end{array}.\nonumber\\
\end{eqnarray}
Thus, for any number of qubits at least $25 \%$ noise is
tolerated.

\subsection{Witnesses for graph states}

\begin{figure}
%\narrowtext
\centerline{\epsfxsize=\columnwidth\epsffile{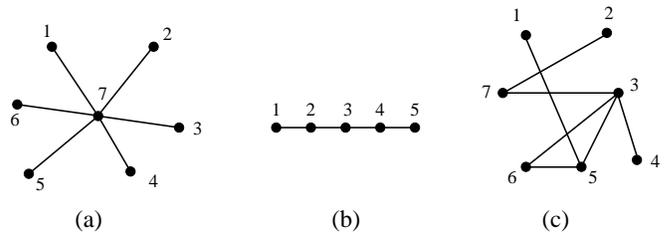}}
\caption{Graphs corresponding to different graph states. (a)
Star graph and (b) cluster state graph. A graph state corresponding
to a star graph is equivalent to a GHZ state under local
unitaries. (c) A seven-vertex graph which has a triangle. Due to
this it is not a two-colorable graph.}
\label{fig_graph}
\end{figure}

Results similar to the ones derived before hold also for graph
states \cite{HE04,DA03}. These states are defined in the following
way: One takes a graph, i.e., a set of $N$ vertices and some edges
connecting them. Edges of this graph are described by the adjacency
matrix $\Gamma$. $\Gamma_{kl}=1$ $(0)$ if the vertices $k$ and $l$
are connected (not connected).
Based on that one can define the stabilizing
operators
\begin{eqnarray}
S_k^{(G_N)}:=
X^{(k)} \prod_{\stackrel{{\rm Neighbors}}{l \mbox{  \tiny{of}  } k}} Z^{(l)}
 =
X^{(k)} \prod_{l \ne k} (Z^{(l)})^{\Gamma_{kl}}.
\label{graph}
\end{eqnarray}
Then, the graph state $\ket{G_N}$ is defined as the $N$-qubit
state fulfilling $\ket{G_N}=S_k^{(G_N)}\ket{G_N}.$ Physically, the
graph provides also a possible generation method: The
graph state can be created from a fully separable state by using
Ising interactions between the connected qubits. In fact, many
useful multi-qubit states can be treated in the graph state formalism,
for instance also GHZ states and cluster states. The corresponding
graphs are shown in Fig.~\ref{fig_graph}.
\\
{\it
{\bf Theorem 7.}
A witness detecting biseparable entanglement close to graph states
can be given as
}
\begin{equation}
\WW^{(G_N)}_{kl} := \eins-S_k^{(G_N)} - S_{l}^{(G_N)},
\label{fsepG}
\end{equation}
{\it
where the spins $(k)$ and $(l)$ are neighbors,
and a witness detecting genuine $N$-party entanglement can be
defined as
}
\begin{equation}
\WW^{(G_N)} := (N-1)\eins - \sum_k S_k^{(G_N)}. \label{wclN}
\end{equation} 
{\it Proof.} The proofs are
essentially the same as before. First one has 
to show that $\eins/2-\ketbra{G_N}$ is a witness for true multipartite
entanglement. Then one can prove that witnesses Eq.~(\ref{fsepG}) and 
Eq.~(\ref{wclN}) detect also only
genuine multipartite entanglement. $\qed$

Now the number of settings needed for measuring $\WW^{(G_N)}$
depends on the graph corresponding to the graph state to be
detected. For this we need the notion of colorability of graphs.
A graph is $M$-colorable, if one can divide the vertices into
$M$ groups and assign to the vertices of each group a color, such
that neighboring vertices have different colors (see Fig.~2).
For two-colorable graphs, only two settings are needed.
In this case
the $S_k^{(G_N)}$ operators can be divided into two groups
corresponding to the two settings. In general, for $M$-colorable
graphs $M$ settings are needed for measuring the witness
$\WW^{(G_N)}$. In this senses, the  most settings ($M=N$) are
needed for the complete graph \cite{BRIEGEL}.

It is, however, important to note at this point that
the colorability of the graph is not an intrinsic and
physical property of the graph state. Usually, a graph state
can be represented by different graphs up to local unitaries,
i.e., different graphs $G_1$ and $G_2$ can result in two graph
states $\ket{G_1}$ and $\ket{G_2}$ which are the same up to
a local change of the basis. Here, $G_1$ and $G_2$ may have
different colorability properties, e.g., $G_1$ may be
two-colorable and $G_2$ $N$-colorable. The question,
which graphs give the same graph state is still an  open
and challenging problem in stabilizer theory. Recently,
much progress has been achieved concerning a subclass
of local unitary transformations, the so-called local
Clifford transformations \cite{Nest}. Efficient algorithms
have been developed which allow the determination of all
graphs which are equivalent within this subclass of local
unitary transformations. These methods can readily be used
to find witnesses with a small number of measurement settings.

\subsection{Obtaining the fidelity of the prepared state}

Let us say that in an experiment we intend to create a GHZ state.
Beside knowing that the prepared state
$\rho$ is entangled, we would also like to know how good its
fidelity is. The fidelity could be measured by measuring the
projector on the GHZ state
\begin{equation}
F:=Tr(\ketbra{GHZ_N}\rho).
\end{equation}
However, we would encounter the same problem as with witnesses:
The number of local settings needed for measuring the projector
increases rapidly with the size.

Fortunately it is possible to obtain a lower bound on the fidelity
from the expectation value of our witnesses. For example, for our
GHZ witness defined in Theorem 3 we have $\WGHZN-2\WPGHZN\ge 0.$
Hence
\begin{equation}
\ketbra{GHZ_N} \ge \frac{1}{2} -\frac{1}{2} \WGHZN,
\end{equation}
where $\WGHZN$ is defined in \EQ{GHZNbetter}. Now a lower bound on
the fidelity $F:=Tr(\ketbra{GHZ_N}\rho)$ can be obtained as
$F':=Tr(P'\rho)$ where
\begin{eqnarray}
P'&:=&\frac{1}{2} -\frac{1}{2} \WGHZN \nonumber\\&=&
\frac{S_1^{(GHZ_N)}+\eins}{2}+\prod_{k=2}^N
\frac{S_k^{(GHZ_N)}+\eins}{2}-\eins. \label{P2GHZ}
\end{eqnarray}
Note that for measuring $P'$ only two local measurement settings are needed.

Let us see how good this lower bound is for the noisy GHZ state
$\vr(p_{\rm noise})$ defined in \EQ{plimit2}. For this state we
have
\begin{eqnarray}
F(p_{\rm noise})&=&1-p_{\rm noise}(1-2^{-N}),\nonumber\\
F'(p_{\rm noise})&=&1-p_{\rm noise}(3/2-2^{-(N-1)}).
\end{eqnarray}
The difference is $\Delta F\approx p_{\rm noise}/2$ for large $N.$
Bounds can be obtained similarly for the fidelity with respect to the cluster state based
on
\begin{equation}
\ketbra{C_N} \ge \frac{1}{2} -\frac{1}{2} \WCN.
\end{equation}

\subsection{Witnesses for mixed states}

In an experiment, after detecting entanglement in the prepared
state, one might be interested to measure and drop one of the
qubits and investigate the state obtained this way. Here we assume
that we do not know the measurement result thus the system is in a
mixed state corresponding to the reduced density matrix of the
remaining qubits. We will show with an example that witnesses can
easily be derived also for this mixed state.

Let us consider a concrete example, namely, measuring a qubit of
the four-qubit cluster state:
\\
{\it
{\bf Example 1.} The witness
}
\begin{eqnarray}
\WW^{(\vr_3)}:=\eins- Z^{(1)}Z^{(2)}- X^{(1)}X^{(2)}Z^{(3)}.
\label{Wrho3}
\end{eqnarray}
{\it
detects entanglement around the state $\vr_3$ which is obtained
from the four-qubit cluster state
\begin{equation}
\ket{C_4'}:=\ket{0000}+\ket{0011}+\ket{1100}-\ket{1111}.
\end{equation}
after the fourth qubit is measured. Note that for \ket{C_4'} we
used a local transformation in order to be able to present the
cluster state in a convenient form in the $Z$ basis.
}
\\
{\it Proof.}  After measuring the fourth qubit, the three-qubit
mixed state is
\begin{eqnarray}
\vr_3&:=&\frac{1}{2}\big(\ketbra{\xi^+}+\ketbra{\xi^-}\big),\nonumber\\
\ket{\xi^+}&:=&\frac{1}{\sqrt{2}}(\ket{00}+\ket{11})\otimes\ket{0},\nonumber\\
\ket{\xi^-}&:=&\frac{1}{\sqrt{2}}(\ket{00}-\ket{11})\otimes\ket{1}.
\label{rho3}
\end{eqnarray}
Note that $\vr_3$ is biseparable, i.e., the third qubit is
unentangled from the first two.

Now we will determine the stabilizer of $\vr_3$. The stabilizers
of states $\ket{\xi^\pm}$, given with their generators, are
$\mathcal{S}^{(\xi^\pm)}:=\GEN{Z^{(1)}Z^{(2)},
X^{(1)}X^{(2)}Z^{(3)},\mp Z^{(3)}}$. We would like to detect
entangled states close to a particular mixture of these two
states. Any mixture of states $\ket{\xi^\pm}$ is stabilized by the
common elements of $\mathcal{S}^{(\xi^\pm)}$: The stabilizer
of these mixed states is
$\mathcal{S}^{(\vr_3)}=\GEN{Z^{(1)}Z^{(2)},
X^{(1)}X^{(2)}Z^{(3)}}$.

Based on that, a stabilizer witness \EQ{Wrho3} can now be
constructed. The constant in \EQ{Wrho3} was determined such that
$\WW^{(\vr_3)}$ detects indeed entangled states only. This can be
proved similarly to the derivation of Theorem 1. Witness
$\WW^{(\vr_3)}$ needs two measurement settings and tolerates noise
if $p_{\rm noise}<1/2.$ $\qed$

\subsection{Entanglement witnesses with nonlocal stabilizing operators}

There are states which do not fit the stabilizer formalism,
however, it is still possible to find a simpler witness than the
one obtained by decomposing the projector. As an example let us
look at the W state
$\ket{W_3}=(\ket{100}+\ket{010}+\ket{001})/\sqrt{3}$. For this
state, the projector-based witness is known to be
\begin{equation}
\WPTHREE=\frac{2}{3}\eins-\ket{W_3}\bra{W_3}.
\end{equation}
It tolerates noise if $p_{\rm noise}<8/21\approx0.38$ and measuring a local
decomposition of this witness requires five measurement settings
\cite{GH03}. In this section we will present witnesses for the
$\ket{W_3}$ state which need only three and two measurement
settings, respectively.

What is new with the stabilizing operators of $\ket{W_3}$?
Clearly, we have to leave the requirement that these operators
should be a tensor product of single-qubit operators. Now the
stabilizing operators must be allowed to be the sum of several
such locally measurable terms.

A set of stabilizing operators with simple local decomposition can
be found in the following way. Let us assume that we create the
$\ket{W_3}$ state from state $\ket{000}$ using a unitary dynamics
$U$. The generators of the stabilizer for $\ket{000}$ are
\begin{equation}
S_k^{(\ket{000})}:=Z^{(k)} \;\;\; \mbox{ for } k=1,2,3.
\end{equation}
Hence
one can get the generators for a group of operators which
stabilize $\ket{W_3}$
\begin{equation}
 S_k^{(W_3)}=U S_k^{\ket{000}}U^\dagger.
\label{gU}
\end{equation}

Let us try to find $U$. It must fulfill
$\ket{W_3}=U\ket{000}$, i.e.,  written in the $Z^{(k)}$ product basis
we must have
\begin{equation}
\frac{1}{\sqrt{3}}
\left[
\begin{tabular}{l}
0 \\ 1 \\ 1 \\0 \\ 1 \\ 0 \\ 0 \\0
\end{tabular}
\right]
=[u_1 u_2 ... u_8]
\left[
\begin{tabular}{l}
1 \\ 0 \\ 0 \\0 \\ 0 \\ 0 \\ 0 \\0
\end{tabular}
\right] \label{mateq}
\end{equation}
Hence, $u_1$ can be obtained. The other
columns of $U$ are not determined by \EQ{mateq} and constrained
only by requiring that $U$ is unitary (i.e., the columns of the
matrix must be orthonormal to each other). Thus $U$ is not unique.
A possible choice for an $U$ fulfilling \EQ{mateq} is
\begin{equation}
U:=\frac{1}{\sqrt{3}}(X^{(1)}Z^{(2)}+X^{(2)}Z^{(3)}+Z^{(1)}X^{(3)}).
\end{equation}

The generators of a group of stabilizing operators can be obtained
based on \EQ{gU}
\begin{eqnarray}
S_1^{(W_3)}&:=&\frac{1}{3}\big(Z^{(1)}+2Y^{(1)}Y^{(2)}Z^{(3)}
+2X^{(1)}Z^{(2)}X^{(3)}\big),\nonumber\\
S_2^{(W_3)}&:=&\frac{1}{3}\big(Z^{(2)}+2Z^{(1)}Y^{(2)}Y^{(3)}+
2X^{(1)}X^{(2)}Z^{(3)}
\big),\nonumber\\
S_3^{(W_3)}&:=&\frac{1}{3}\big(Z^{(3)}+2Y^{(1)}Z^{(2)}Y^{(3)}
+2Z^{(1)}X^{(2)}X^{(3)}\big).\nonumber\\
\end{eqnarray}
These three stabilizing operators uniquely define the $W$
state. Again, by multiplying them with each other, other
stabilizing operators can be found. However, with the exception of
$S_1^{(W_3)}S_2^{(W_3)}S_3^{(W_3)}=-Z^{(1)}Z^{(2)}Z^{(3)}$, they
are all nonlocal.

Now let us try to create an entanglement witness which detects
genuine multi-party entanglement, but is still easier to measure
than the projector-based witness $\WPTHREE$. Consider the
following witness constructed from some of the elements of the
group generated by $S_k^{(W_3)}$
\begin{eqnarray}
\WW&=&c_0\eins-S_1^{(W_3)}S_2^{(W_3)}-S_2^{(W_3)}S_3^{(W_3)}
-S_1^{(W_3)}S_3^{(W_3)}\nonumber\\
&-&c_1S_1^{(W_3)}S_2^{(W_3)}S_3^{(W_3)},
\end{eqnarray}
where $c_0$ and $c_1$ are positive constants.
The expectation value of $\WW$ is minimal for the W state. Constants
$c_0$ and $c_1$ must be determined such that for some $\alpha>0$ we have
$\WW-\alpha\WPTHREE\ge 0$, and also we have the possible
best noise tolerance. Thus we have:
\\
{\it
{\bf Theorem 8.}
A witness detecting genuine three-qubit entanglement around
a $\ket{W_3}$-state is
}
\begin{eqnarray}
&&\WTHREE:= \frac{11}{3} \eins+
2Z^{(1)}Z^{(2)}Z^{(3)}\nonumber\\&&\;\;\;-\frac{1}{3}\sum_{k\ne
l}\big(2X^{(k)}X^{(l)}+2Y^{(k)}Y^{(l)}-Z^{(k)}Z^{(l)}\big) .
\end{eqnarray}
{\it
It requires three measurement settings and tolerates noise if
$p_{\rm noise}<4/15\approx 0.27$.

One can
also leave out the $Z^{(k)}$ terms:  The following entanglement witness
detects genuine three-qubit entanglement close to a $\ket{W_3}$-state}

\begin{eqnarray}
{\WW}' := (1+\sqrt{5})\eins &-&
X^{(1)}X^{(2)}-X^{(2)}X^{(3)}-X^{(1)}X^{(3)}\nonumber\\
&-&Y^{(1)}Y^{(2)}-Y^{(2)}Y^{(3)}-Y^{(1)}Y^{(3)}. \label{WN}
\nonumber\\
\end{eqnarray}
{\it This witness requires the
measurement of two settings and tolerates noise if $p_{\rm
noise}<(3-\sqrt{5})/4\approx0.19.$ The proof is given in Appendix C.}

\section{Criteria using variances and uncertainty relations}

Let us now explain, how criteria in terms of variances can be
derived, using uncertainty relations \cite{nonlin, lu1, eu1}.
In Ref.~\cite{lu1} the following recipe was presented for the
derivation of such criteria, called the {\it local uncertainty relations}
(LURs).
Consider a bipartite quantum system and let $A_i$ be some observables
on one party, fulfilling a bound
\begin{equation}
\sum_i \delta^2(A_i) \geq U_A
\label{lur1}
\end{equation}
for all states on this party. Here,
$\delta^2(A_i)=\mean{A^2}-\mean{A}^2$ denotes the variance of the
state. This inequality is an uncertainty relation for the $A_i$, with
$U_A > 0$ iff the observables $A_i$ have no common eigenstates. Let
us  assume that we have also observables $B_i$ on the second
system, fulfilling a similar bound $\sum_i \delta^2(B_i) \geq
U_B.$ Then, we may look at the observables $M_i=A_i\otimes \eins +
\eins \otimes B_i$ on the composite system. As it was shown in
Ref. \cite{lu1} for separable states
\begin{equation}
\sum_i \delta^2(M_i) \geq U_A +U_B
\end{equation}
has to hold, and a violation of this bound implies that the state
is entangled.

Now we show how the witness $\WW_m^{(GHZ_N)}$ in
Eq.~(\ref{fsepGHZ}) for GHZ states can be improved using nonlinear
terms:
\\
{\it
{\bf Theorem 9.} Let us define
}
\begin{eqnarray}
A_1&=&X^{(1)} X^{(2)} \cdot \cdot \cdot X^{(k)}  ,
\nonumber\\
A_2&=&Z^{(k)},
\nonumber\\
B_1&=&-X^{(k+1)} X^{(k+2)} \cdot \cdot \cdot X^{(N)}  ,
\nonumber\\
B_2&=&-Z^{(k+1)},
\end{eqnarray}
{\it
for $k=1,2,...,N-1.$ Using these operators, the following necessary
condition for separability can be given
}
\begin{eqnarray}
1 &-& \mean{S_1^{(GHZ_N)}} - \mean{S_{k+1}^{(GHZ_N)}} \nonumber\\&-&
\frac{1}{2}\big( \mean{A_1 + B_1}^2 + \mean{A_2 +  B_2}^2 \big)
\geq 0. \label{nonlincrit}
\end{eqnarray}
{\it Proof.} For the uncertainties of observables $A_k$ and $B_k$, one has the bounds
\bea \delta^2(A_1)+\delta^2(A_2) \geq 1,
\nonumber\\
\delta^2(B_1)+\delta^2(B_2) \geq 1. \label{uncert} \eea Knowing
that $\mean{X^{(k)}}^2+\mean{Z^{(k)}}^2 \leq 1,$ which implies
that $\delta^2(X^{(k)})+\delta^2(Z^{(k)}) \geq 1$ these bounds
should not be a surprise. A detailed proof which relies on the
fact that $A_1$ and $A_2$ anti-commute, is given in Appendix C.
From this and the method of the LURs \EQ{nonlincrit} follows.
$\qed$

Eq.~(\ref{nonlincrit}) is a nonlinear necessary condition for
separability. It can be considered as a nonlinear "refinement" of
$\WGHZN_m$ in Eq.~(\ref{fsepGHZ}), which improves the witness by
subtracting the squares of some expectation values. The fact that
LURs can sometimes improve entanglement witnesses was, for
bipartite systems,  first observed in Ref.~\cite{lu2}. There, the
magnitude of the improvement for a special case was also
investigated numerically. In the given case for GHZ states it is
important to note that the LUR does not improve the noise
tolerance, when the GHZ state is mixed with white noise. This is
due to the fact that for the totally mixed state as well as for
the GHZ state the squared mean values in Eq.~(\ref{nonlincrit})
vanish. However, there are many states which are not of this form
and which are detected by the LUR and not by the witness.

In fact, many of the witnesses from the previous sections can be
improved via the method presented above.  For instance, one may
add the extra observables $A_3=X^{(1)}X^{(2)}\cdot\cdot\cdot
X^{(k-1)}Y^{(k)}, B_3=-Y^{(k+1)}X^{(k+2)}\cdot\cdot\cdot X^{(N)}$
to the observables $A_1, ...,B_2$ from above. Then
$\delta^2(A_1)+\delta^2(A_2)+\delta^2(A_3) \geq 2$ and the same
bound holds for the $B_i.$ This leads to the separability
criterion \bea 1 &-& \mean{S_1^{(GHZ_N)}}-\mean{S_m^{(GHZ_N)}} -
\mean{S_1^{(GHZ_N)}S_{m}^{(GHZ_N)}}-
\nonumber \\
&-&\frac{1}{2}
\big(
\mean{A_1 + B_1}^2
+
\mean{A_2 +  B_2}^2
+
\mean{A_3 +  B_3}^2
\big)
\geq 0
\nonumber \\
\eea
which improves the witness of Eq.~(\ref{fsepGHZprime}). Also the witness in
Eq.~(\ref{fsepC}) can be improved using the same methods, leading to the
separability condition
\bea
1&-&\mean{S_k^{(C_N)}}-\mean{S_{k+1}^{(C_N)}}
- \frac{1}{2}
\mean{Z^{(k-1)}X^{(k)}-Z^{(k+1)}}^2
-
\nonumber \\
&&
- \frac{1}{2} \mean{Z^{(k)} - X^{(k+1)}Z^{(k+2)}}^2 \geq 0,
\eea
and the witness for graph states can be improved in a similar
manner.

To conclude, it turned out that LURs can improve the
presented witnesses which were designed for ruling out full
separability. The witnesses for the detection of genuine
multipartite entanglement can not be so simply improved, mainly for
two reasons. First, LURs are specially designed for bipartite
systems, and no generalization to multipartite systems in known
so far. Second, according to  our definition of multipartite
entanglement a state which is a convex combination of states which
are biseparable with respect to different partitions is also
biseparable. This implies that it is not enough to show that a
state is biseparable with respect to each partition in order to
show that it is multipartite entangled. However, if one defines a
state to be multipartite entangled if it is not biseparable with
respect to any partition (as it is done sometimes in the
literature, see, e.g., Ref.~\cite{lu3}) then LURs can be used to
detect multipartite entanglement by ruling out biseparability for
every bipartite split (as proposed in \cite{lu1}).

\section{Criteria using entropic uncertainty relations}

Let us now discuss another possibility of deriving separability
criteria in terms of uncertainty relations, namely criteria based
on entropic uncertainty relations. As we will see, the stabilizer
formalism allows us to formulate easily such criteria, however,
they are not as strong as the witnesses or the variance based
criteria.

The recipe to derive such criteria was described in
Ref.~\cite{eu1} and goes as follows: If we have an observable $M$
we can define $\PP(M)_\varrho=(p_1, ..., p_n)$ as the probability
distribution of the different outcomes for measuring $M$
in the state $\varrho.$  One can characterize the uncertainty of this
measurement by taking the entropy of this probability
distribution, i.e., by defining $H(M):=H(\PP(M)_\varrho).$ Here,
we only consider the entropy to be the standard Shannon entropy
$H(\PP):= - \sum_k p_k \ln(p_k),$ however, a more general entropy
like the Tsallis entropy may also be used. Given two observables $M$
and $N$ which do not share a common eigenstate, there must exist a
strictly positive constant $C$ such that $ H(M)+H(N) \geq C $
holds. The difficulty in this so-called entropic uncertainty
relation (EURs) lies in the determination of $C.$ For results on
this problem see Refs. \cite{eu2,eu3}. For the detection of
entanglement, the following result has been proved \cite{eu1}:
Let $A_1,A_2$ and $B_1,B_2$ be observables with nonzero
eigenvalues on Alice's (respectively, Bob's) space obeying an EUR
of the type
\begin{equation}
H(A_1)+H(A_2)\geq C
\end{equation}
and the same bound for $B_1,B_2.$ If $\varrho$ is
separable, then
\begin{equation}
H(A_1 \otimes B_1)_\varrho +
H(A_2 \otimes B_2)_\varrho
\geq C.
\end{equation}
holds. For entangled states this bound can be violated, since
$A_1\otimes B_1$ and $A_2\otimes B_2$ might be degenerate and have
a common entangled eigenstate.

In order to apply this scheme to the detection of entanglement in
the stabilizer formalism we have to recall some more facts. Assume
that we have two observables of the form $M=\sum \mu_i P_i$
and $N=\sum_j \nu_j Q_j$ where the $P_i, Q_j$ are the projectors
on the eigenspaces. Here, we do not require $M$ and $N$ to be
non-degenerate, i.e., the $P_i$ and $Q_j$ may have ranks larger
than one.  In this situation, it was shown in Ref.~\cite{eu3} that
for these observables the EUR
\begin{equation}
H(M)+H(N)\geq - \ln
\max_{ij} Z_{ij}
\label{eur1}
\end{equation}
holds, where $Z_{ij} = \Vert
P_i Q_j \Vert = \max_{\ket{\psi}}\sqrt{\bra{\psi}(P_i Q_j)^\dagger
(P_i Q_j)\ket{\psi}}$ is the norm of the operator $P_i Q_j.$ This
has two consequences. First, it follows immediately that for one
qubit
\begin{equation}
H(X) + H(Y) \geq \ln(2),
\label{eur2}
\end{equation}
holds, and similar relations hold for $Y,Z$ etc. Second, if $A$ is
an observable consisting of Pauli measurements on $N$ qubits, e.g.,
$A=X^{(1)}Y^{(2)}...Z^{(N)},$ then for $N+1$ qubits the EUR
\begin{equation}
H(A \otimes X^{(N+1)}) + H(\eins_{2^N} \otimes Y^{(N+1)})
\geq \ln(2), \label{eur3}
\end{equation}
holds. Here $\eins_{2^N}$ denotes
the identity on the first $N$ qubits. Similar result hold
also, if the observables on the qubit $N+1$ are replaced by other
Pauli matrices. Eq.~(\ref{eur3}) can directly be proved from
Eq.~(\ref{eur1}) by identifying the corresponding $P_i$ and $Q_j.$

Armed with these insights, we can formulate now entropic criteria
for stabilizer states:
\\
{\it
{\bf Theorem 10.} For GHZ and cluster
states, respectively, the following necessary conditions for
biseparability can be given using entropic uncertainties
\begin{eqnarray}
\sum_{k=1}^N H(S_k^{(GHZ_N)}) &\geq& \ln(2), \label{eur4}\\
\sum_{k=1}^N H(S_k^{(C_N)}) &\geq& \ln(2). \label{eur5}
\end{eqnarray}
Any state violating these conditions is genuine $N$-qubit
entangled. Note that for GHZ and cluster states the left hand side
of Eq.~(\ref{eur4}) and Eq.~(\ref{eur5}), respectively, is zero.
}
\\
{\it Proof.} To prove Eq.~(\ref{eur4}) it suffices to
look at pure biseparable states, since the entropy is concave in
the state. So let us assume that $\ket{\psi}=\ket{a}\ket{b}$ is a
biseparable state. For definiteness, we assume that $\ket{a}$ is a
state of the the qubits $1,2,...,k$ and $\ket{b}$ is a state of
the qubits $k+1,k+2,...,N;$ the proof of the other cases is
similar. In order to apply the recipe from above, we have to show
that on the first $k$ qubits the EUR
\begin{equation} H(X^{(1)}... X^{(k)}) +
\sum_{i=2}^k H(S_k^{(GHZ_N)}) + H(Z^{(k)}) \geq \ln(2)
\end{equation}
holds. This is easy to see, since $H(X^{(1)}\cdot\cdot\cdot
X^{(k)}) + H(Z^{(k)}) \geq \ln(2)$ is valid, due to
Eq.~(\ref{eur3}). Similar ideas can be used for cluster states.
The proof is essentially the same as for the GHZ state. Similar
statements can also be derived for arbitrary graph states. $\qed$

As we have seen, it is quite straightforward to formulate entropic
criteria for stabilizer states. However, one should clearly state
that the criteria in the presented form are not very strong. For
instance, the criterion in Eq.~(\ref{eur4}) detects states mixed
with white noise only for $p_{\rm noise}<0.123$ for $N=3$ and
$p_{\rm noise}<0.083$ for $N=4.$ The robustness to noise can, as
shown for some two-qubit cases in Ref. \cite{eu1}, be improved, if
other entropies than the Shannon entropy are used. However, for
these entropies no such general bound as in Eq.~(\ref{eur1}) is
known. So only a better understanding of the EURs can help to
explore the full power of the entropic criteria.

\section{Conclusions}

In summary, we have shown that stabilizer theory can be used very
efficiently to derive sufficient criteria for entanglement.
Knowing  some stabilizing operators of a state allows for an easy
derivation of a plethora of entanglement criteria which detect
states in the vicinity of the state. This holds for linear as well
as nonlinear criteria, and for the different types of entanglement
in the multipartite setting. We also noted that the resulting
criteria exhibit several interesting features: They all are easy
to implement in experiments, some of them have interesting
connections to Bell inequalities, others are nonlinear
improvements of witnesses.

A natural continuation of the present work lies in the systematic
extension of the presented ideas to states which do not fit
directly in the stabilizer formalism. GHZ, cluster and graph
states are not the only multipartite states which are interesting
from the viewpoint of quantum information science. As we have
shown in the example of the W state, also for states outside the
stabilizer formalism similar ideas can be applied by identifying
their nonlocal stabilizing operators. Exploring this direction
might help to clarify the structure of multipartite entanglement.

\section{Acknowledgment}

We thank M.~Aspelmeyer, H.J.~Briegel, D.~Bru{\ss}, {\v
C}.~Brukner, J.I.~Cirac, T.~Cubitt, J.~Eisert,
J.J.~Garc\'{\i}a-Ripoll, P.~Hyllus, M.~Lewenstein,
M.~Van~den~Nest, A.~Sanpera, M.~Seevinck, J.~Uffink, M.M.~Wolf,
and M.~\.Zukowski for useful discussions. We also acknowledge the
support of the DFG (Graduierten\-kolleg 282), the Marie Curie
Individual Fellowship of the European Union (Grant No.
MEIF-CT-2003-500183), the EU projects RESQ and QUPRODIS, and the
Kom\-pe\-tenz\-netz\-werk
Quan\-ten\-in\-for\-ma\-tions\-ver\-ar\-bei\-tung der
Bayerisch\-en Staats\-re\-gie\-rung.

\section*{APPENDIX A: The stabilizer group}

We summarize the properties of the stabilizing operators of a
given $N$-qubit quantum state $\ket{\Psi}$. These are locally
measurable operators $\widetilde{S}$ for which
$\widetilde{S}\ket{\Psi}=\ket{\Psi}$. They have $\pm 1$
eigenvalues. For states considered in this paper (GHZ, cluster and
graph states) they are tensor products of Pauli spin matrices.

Let us now list the properties of the set $\mathcal{S}$ containing
the stabilizing operators of $\Psi$. For any
$\widetilde{S}_1,\widetilde{S}_2,\widetilde{S} \in \mathcal{S}$
\begin{eqnarray}
\widetilde{S}_1\widetilde{S}_2&\in& \mathcal{S},\nonumber\\
{[}\widetilde{S}_1,\widetilde{S}_2]&=&0,\nonumber\\
\widetilde{S}^2&=&\eins.
\end{eqnarray}
Now it is clear that the elements of
$\mathcal{S}$ form a commutative (Abelian) group. It is called the
stabilizer. One of its $2^N$ elements is given as
\begin{equation}
\widetilde{S}_k := \prod_{l=1}^N (S_l)^{\alpha_{kl}}, \label{defS}
\end{equation}
where $\alpha_{kl}$ is the $l$th digit of the binary number (i.e.,
$N$-tuple of $\{0,1\}$) corresponding to the number
$k\in\{0,1,2,...,2^{N}-1\}.$ Operators $\{S_l\}_{l=1}^N$ are the
generators of the group. If $\alpha_{kl}=1$ then we say that
$\widetilde{S}_k$ {\it contains} generator $S_l$ in its definition
of the type \EQ{defS}.

\section*{APPENDIX B: Proof of optimality for
the GHZ and cluster state witnesses }

In this section we will prove that the GHZ and cluster state
witnesses defined in Eq. (\ref{GHZNbetter}) and Eq. (\ref{CN}),
respectively, are optimal. That is, it is not possible to find a
stabilizer witness $\WW$ which needs only two measurement
settings, has the property that for some $\alpha> 0$ we have
$\WW-\alpha \widetilde{\WW}\ge 0$ and has better noise tolerance
than the witnesses presented in this paper. Here $\widetilde{\WW}$
denotes a projector-based witness.

Before presenting the proof, let us analyze what we understand on
measurement settings. Let us consider operators which can be
measured with one measurement setting. These form a group, members
of which commute {\it locally} (for an explanation of local
commuting see Sec.~II.B). The projector cannot be measured with
two settings, since the stabilizing operators cannot be divided
into two locally commuting subgroups. So we will find two
subgroups, which contain as many stabilizing operators as
possible. Let us consider GHZ states first. The largest such
subgroup is $\GEN
{S^{(GHZ_N)}_2,S^{(GHZ_N)}_3,...,S^{(GHZ_N)}_N}$. Here the group
is given by the generators. The operators in this group commute
locally since they all contain only $Z^{(k)}$'s, and neither
$X^{(k)}$'s nor $Y^{(k)}$'s. All the other elements of the
stabilizer have the form $S^{(GHZ_N)}_1 Q$ where $Q\in
\GEN{S^{(GHZ_N)}_2,S^{(GHZ_N)}_3,...,S^{(GHZ_N)}_N}$. All these
operators contain only $X^{(k)}$'s and $Y^{(k)}$'s, and do not
contain $Z^{(k)}$'s. None of these commute locally with any other
element of the stabilizer, except with the identity. Thus the
other subgroup can contain only one such operator and the
identity. 

Now the two subgroups corresponding to the two local settings are
\begin{eqnarray}
\mathcal{L}^{(GHZ_N)}_1&=&\{S^{(GHZ_N)}_1Q,\eins\},\nonumber\\
\mathcal{L}^{(GHZ_N)}_2&=&\GEN
{S^{(GHZ_N)}_2,S^{(GHZ_N)}_3,...,S^{(GHZ_N)}_N}.
\end{eqnarray}
Any operator which is a linear combination of operators in
$\mathcal{L}_1$ can be measured by the first measurement setting.
Similarly, any operator which is linear combination of operators
in $\mathcal{L}_2$ can be measured by the second measurement
setting. It is an important question whether
there is another choice for $\mathcal{L}_1$ and $\mathcal{L}_2$. In
order to answer this, we have to remember that we want to detect
genuine $N$-qubit entanglement. Thus we must find a witness for
which $\exs{\WW}$ has a unique minimum for GHZ states. For that,
we must be able to measure a complete set of generators with the
two settings. (For more details, see Sec. II.D. ) It can be seen
that no other two subgroups can be found which fulfill this
requirement. For simplicity, in the following we choose $Q=\eins.$ 
The two measurement settings corresponding to this case are shown in
Fig.~\ref{fig_settings}(a).
Choosing $Q$ to be
not the identity would not change our argument and would not lead
to witnesses with better noise tolerance than the ones presented
here. 

As discussed before, the eigenvectors of the generators
$\{S_k^{(GHZ_N)}\}_{k=1}^N$ of the stabilizer form a complete
basis (GHZ basis). We will use it to represent states of the
$N$-qubit Hilbert space. Let us use $N$-tuples of $\{0,1\}$ for
labeling the basis states. If the $k$th digit is 0 (1) then for
the basis state $\exs{S_k^{(GHZ_N)}}=+1$
($\exs{S_k^{(GHZ_N)}}=-1$). Now $\ket{[00...0]}$ is the GHZ state.
Here square brackets are used in order to draw our attention to
the fact that the GHZ basis is not a product basis and the $N$
digits do not correspond to a physical partitioning of the system.

Next, let us use this labeling to order the $2^N$ basis states
from $\ket{[00...0]}$ to $\ket{[11...1]}$. What is the matrix form
of $S_1^{(GHZ_N)}$ in this basis? It is clearly diagonal.
Moreover, it must give $+1$ and $-1$ expectation values for states
of the form $\ket{[0 s_2 s_3 ... s_N ]}$ and
$\ket{[1 s_2 s_3 ... s_N ]}$, respectively.
Knowing this, it must have the form
${diag}(+1 , -1)\otimes \eins_{2^{N-1}}.$ Here ${diag}$ denotes a
diagonal matrix and the size of the identity is indicated by a
subscript. Note again that the tensor product does not correspond
to a physical partitioning of the system. However, the matrix form
of $S_1^{(GHZ_N)}$ in the GHZ basis is the same as the matrix form
of $Z^{(1)}$ in the product basis. Similarly, the matrix form of
$S_k^{(GHZ_N)}$ in the GHZ basis is the same as the matrix form of
$Z^{(k)}$ in the product basis.

After these considerations about measurement settings, let us
start our proof. Let us find out how operators measurable by the
first setting look like in the GHZ basis. 
Now it is clear that such operators must
have the form $A\otimes\eins_{d_2}$ where matrix $A$ is of size
$d_1=2$ and $d_2=2^{(N-1)}$. Operators measurable by the second
setting have the form $\eins_{d_1}\otimes B$ in this basis. Here
$B$ is of size $d_2$. We construct our witness from two operators
corresponding to the two measurement settings as
\begin{equation}
\WW:=c\eins_{d_1}\otimes\eins_{d_2}
-A\otimes\eins_{d_2}-\eins_{d_1}\otimes B,
\end{equation}
where $c$ is a constant. Now we will find an operator $\WW$ for
which for some $\alpha>0$ we have
$\WW-\alpha\widetilde{\WW}^{(GHZ_N)}\ge 0$ and which is optimal
from the point of view of noise tolerance. Without the loss of
generality, we set $\alpha=2$. Both $\WW$ and the projector
witness $\widetilde{\WW}^{(GHZ_N)}$ are diagonal in the GHZ basis.
Hence from the condition $\WW\ge2\widetilde{\WW}^{(GHZ_N)}$, the
following constraints on the diagonal elements $a_k$ and $b_k$ of
$A$ and $B$, respectively, can be obtained
\begin{eqnarray}
c-a_1-b_1&\ge& -1, \nonumber\\
c-a_k-b_l&\ge& +1;\;\;k \ge 2;\; l\ge 1, \nonumber\\
c-a_k-b_l&\ge& +1;\;\;k \ge 1;\; l\ge 2.
\label{constraints}
\end{eqnarray}
The maximal noise $p_{\rm limit}$ tolerated by our witness can be
computed as given in \EQ{plimit2}. For this formula we have to use
\begin{eqnarray}
\langle GHZ_N|\WW |GHZ_N\rangle&=&c-a_1-b_1,\nonumber\\
2^{-N} Tr( \WW ) &=& c-\frac{1}{d_1}\sum_{k=1}^{d_1} a_k-\frac{1}{d_2}\sum_{k=1}^{d_2}b_k.\nonumber\\
\end{eqnarray}
In order to find the optimal witness, constant $c$ and the
elements of $A$ and $B$ must be chosen such that the noise
tolerance $p_{\rm limit}$ is maximized, under the constraints
Eq.~(\ref{constraints}). This is the case if all the three
inequalities in Eq.~(\ref{constraints}) are saturated. Thus we
obtain $c=3$, $a_1=b_1=2$, $a_k=b_k=0$ for $k\ge 2$ and
\begin{equation}
p_{\rm limit}=\frac{1}{4-2/d_1-2/d_2}.\label{noised1d2}
\end{equation}
The witness
obtained this way coincides with the witness given in
Eq.~(\ref{GHZNbetter}).

Now let us turn to cluster states. We will need the following
lemma:
\\
{\it
{\bf Lemma 1.} Let us consider the following three
subsets of the stabilizer $\mathcal{S}^{(C_N)}$
}
\begin{eqnarray}
\mathcal{P}_1:=\{ A \in \mathcal{S}^{(C_N)}: &&A \text{ contains }
S_l^{(C_N)}\\&&\text{ and does not contain } S_{l+1}^{(C_N)} \},\nonumber\\
\mathcal{P}_2:=\{ A \in \mathcal{S}^{(C_N)}: &&A \text{ contains }
S_{l+1}^{(C_N)}\\ && \text{ and does not contain } S_{l}^{(C_N)} \},\nonumber\\
\mathcal{P}_3:=\{ A \in \mathcal{S}^{(C_N)}: && A \text{ contains
both } S_{l+1}^{(C_N)} \nonumber\\ && \text{ and } S_{l}^{(C_N)} \}.\nonumber\\
\end{eqnarray}
{\it
If a local measurement setting makes it possible to measure an
operator in $\mathcal{P}_n$, then it does not make it possible to
measure any operator in $\mathcal{P}_m$ for $m\ne n.$
}
\\
{\it Proof.} First let us prove our Lemma for $l=1.$
Table~\ref{table1}(a) shows for some elements of the stabilizer
which measurements on qubits $(1)$ and $(2)$ are needed to measure
them. Here for brevity superscript $C_N$ is omitted. A binary
pattern indicates whether a given element of the stabilizer
contain or does not contain generators $S_1$, $S_2,$ and $S_3.$
For example, entry $100$ represents operators which contain $S_1$
and do not contain $S_{2/3}.$ Such operators are, for example,
$S_1$, $S_1S_4$ and $S_1S_4S_6.$ In the second row of the table
"$X$","$Y$" and "$Z$" represent Pauli spin matrices. "$1$"
indicates that no measurement is needed on the given qubit. For
example, in the column of $101$ entry "X1" indicates that for
measuring $S_1S_3=X^{(1)}X^{(3)}Z^{(4)}$ an $X$ measurement is
needed for qubit $(1)$ and no measurements are needed for qubit
$(2).$

The left two columns correspond to operators  of set
$\mathcal{P}_1$ which contain $S_1$ and do not contain $S_2.$ The
middle two columns correspond to operators of set $\mathcal{P}_2$
which contain $S_2$ and do not contain $S_1.$ The right two
columns correspond to operators of set $\mathcal{P}_3$  which
contain both $S_1$ and $S_2.$ Based on the table we see that only
$100$ and $101$ have compatible measurements for the first two
qubits: $"XZ"$ and $"X1"$, respectively. An operator of
$\mathcal{P}_n$ cannot be measured together with any operator of
$\mathcal{P}_m$ for $m \ne n.$ Thus Lemma 1 works for $l=1.$

\begin{table}

\begin{tabular}{|l||l|l||l|l||l|l|}\hline
     $S_1S_2S_3$       & $100$ & $101$ & $010$ & $011$ & $110$ & $111$ \\\hline
     $O^{(1)}O^{(2)}$  & $XZ$  & $X1$  & $ZX$  & $ZY$  & $YY$  & $YX$  \\\hline
\end{tabular}\\

\vskip 0.2cm
\centerline{(a)}
\vskip 0.5cm

\begin{tabular}{|l||l|l|l|l|}\hline
     $S_{k-1}S_{k}S_{k+1}S_{k+2}$  & $0100$ & $0101$ & $1100$ & $1101$   \\\hline
     $O^{(k)}O^{(k+1)}$            & $XZ$   & $X1$   & $YZ$   & $Y1$     \\\hline\hline
     $S_{k-1}S_{k}S_{k+1}S_{k+2}$  & $0010$ & $0011$ & $1010$ & $1011$   \\\hline
     $O^{(k)}O^{(k+1)}$            & $ZX$   & $ZY$   & $1X$   & $1Y$     \\\hline\hline
     $S_{k-1}S_{k}S_{k+1}S_{k+2}$  & $0110$ & $0111$ & $1110$ & $1111$   \\\hline
     $O^{(k)}O^{(k+1)}$            & $YY$   & $YX$   & $XY$   & $XX$     \\\hline
\end{tabular}

\vskip 0.2cm \centerline{(b)} \caption{(a) Local measurements
needed for measuring some elements of the stabilizer on qubits
$(1)$ and $(2)$ for a cluster state. The binary pattern indicates
presence or absence of generators $S_{1},$ $S_{2}$ and $S_{3}.$
(b) Local measurements needed for qubits $k$ and $k+1$ where
$k\ge2.$ The binary pattern indicates presence or absence of
generators $S_{k-1}$, $S_{k}$, $S_{k+1}$ and $S_{k+2}.$ For
details see text.} \label{table1}
\end{table}

Now let us prove, that if our Lemma is true for $l=k-1$, then it
is true for $l=k$. Table~\ref{table1}(b) shows the measurements
for qubits $k$ and $k+1$ for a particular set of operators. An
entry of the table, let us say $0100$, represents all operators
which contain $S_k$, and does not contain $S_{k-1}$,$S_{k+1}$ and
$S_{k+2}.$ Now the measurements for qubits $k$ and $k+1$ for
$0101$ seem to be compatible with the measurements for $1010,$
$1011,$ $1110,$ and $1111.$ However, we assumed that our Lemma is
true for $l=k-1.$ Thus $0101$ and $1010$ cannot be measured with
the same setting. For the same reason, $0101$ and $1011$ cannot be
measured with the same setting. Taking the other operators having
compatible measurements for qubits $k$ and $k+1$, we find that
measuring them with the same setting is prohibited by our Lemma
for $l=k-1$. $\qed$

Lemma 1 limits the size of the subgroup of the stabilizer of
$\ket{C_N}$ measurable by one measurement setting. Namely, for
even $N$ the largest subgroup can be given with $N/2$ generators,
while for odd $N$ the maximum is $(N+1)/2.$ Two subgroups of the
stabilizer allowing for the measurement of a complete set of
generators are
\bea\mathcal{L}_1^{(C_N)}:=\GEN{S^{(C_N)}_1,S^{(C_N)}_3,S^{(C_N)}_5,...},\nonumber\\
\mathcal{L}_2^{(C_N)}:=\GEN{S^{(C_N)}_2,S^{(C_N)}_4,S^{(C_N)}_6,...}.\eea
Now $d_1=d_2=2^{N/2}$ for even $N$ and $d_{1/2}=2^{(N\pm 1)/2}$
for odd $N$. There are other possibilities for the two measurement
settings, however, they do not give a better noise tolerance since
the optimal noise tolerance \EQ{noised1d2} depends only on $d_1$
and $d_2$, and does not depend on which elements of the stabilizer
are measured by the two settings. Optimization leads to witness
(\ref{CN}) constructed using two projectors, corresponding to the
two settings. Thus witness given in Eq. (\ref{CN}) is also
optimal.

\section*{APPENDIX C: Some technical calculations}

{\it Calculations for the W state --- } Here, we prove that
\EQ{WN} describes an entanglement witness detecting states with
three-qubit entanglement around a W state. Note that this witness
was constructed independently from the stabilizer theory. The
proof is also useful in general, since it shows how to find the
minimum of an operator expectation value for biseparable states
analytically.

Let us first assume $(1)(23)$ biseparability. Then for a state of
the form $\Psi=\Psi_1\otimes\Psi_{23}$
\begin{eqnarray} &&\exs{\WW'}=(1+\sqrt{5})-
\exs{X^{(1)}}\exs{X^{(2)}}
-\exs{X^{(1)}}\exs{X^{(3)}}\nonumber\\
&&-\exs{X^{(2)}X^{(3)}} -\exs{Y^{(1)}}\exs{Y^{(2)}}
-\exs{Y^{(1)}}\exs{Y^{(3)}}\nonumber\\
&&-\exs{Y^{(2)}Y^{(3)}} \nonumber\\
&&=(1+\sqrt{5})-\exs{X^{(1)}}_1[\exs{X^{(2)}+X^{(3)}}_{23}]\nonumber\\
&&+\exs{Y^{(1)}}_1[\exs{Y^{(2)}+Y^{(3)}}_{23}]
+[\exs{X^{(2)}X^{(3)}+Y^{(2)}Y^{(3)}}_{23}]\nonumber\\
&&=\exs{F(x,y)}_{23},\label{exsA}
\end{eqnarray}
where
\begin{equation}
x:=\exs{X^{(1)}}_1;\;\;\;\;
y:=\exs{Y^{(1)}}_1.
\end{equation}
Here $\exs{...}_1$ and $\exs{...}_{23}$, respectively,  denote
expectation value computed for $\Psi_1$ and $\Psi_{23}$. To be
explicit, matrix $F$, as the function of two real parameters, is
given
\begin{eqnarray}
F(x,y) &:=&(1+\sqrt{5})\eins-
x[X^{(2)}+X^{(3)}]-
y[Y^{(2)}+Y^{(3)}]\nonumber\\
&-&[X^{(2)}X^{(3)}+Y^{(2)}Y^{(3)}].
\end{eqnarray}
Note that operator $F$ acts on the second and third qubits while
its two parameters depend on the state of the first qubit.
The expectation value of $F(x,y)$ can be bounded
from below
\begin{eqnarray}
\exs{F(x,y)}_{23}\ge\Lambda_{\rm min}[F(x,y)]=
\sqrt{5}-\sqrt{1+4(x^2+y^2)},\nonumber\\
\label{eqF}
\end{eqnarray}
where $\Lambda_{\rm min}(F)$ is the smallest eigenvalue of $F$.
The right hand side of Eq. (\ref{eqF}) was obtained by finding the
matrix eigenvalues analytically. Using
$\exs{X^{(1)}}^2+\exs{Y^{(1)}}^2\le~1$, we obtain that
$\exs{\WW'}$ is bounded from below by
\begin{eqnarray}
\exs{\WW'} &\ge& \min_{x^2+y^2\le 1}\{ \Lambda_{\rm min}[
F(x,y)]\}=0.
\end{eqnarray}
Thus our witness has a non-negative expectation value for
biseparable pure states with partitioning $(1)(23)$. Due to that
$\WW'$ is symmetric under the permutation of qubits, $\ex{\WW'}\ge
0$ holds also for pure biseparable states with a partitioning
$(12)(3)$ and $(13)(2).$ It is easy to see that the bound is also
valid for mixed biseparable states. In contrast, for three-qubit
entangled states we have $\ex{\WW'}\ge \sqrt{5}-3$. Among pure states, 
the equality
holds only for $\ket{W_3}$ and
$\ket{\overline{W_3}}=(\ket{011}+\ket{101}+\ket{110})/\sqrt{3}$,
and their superpositions.

{\it Calculations for the LURs described in Sec. IV--- } To
compute the bounds required for the derivation of the LURs we show
the following. Let $A_i,$ for $i=1,2,...,n$ be some observables,
which anti-commute pairwise (i.e., $A_i A_j + A_j A_i =0$ for all
$i \neq j$) and which have all $\pm 1$ eigenvalues (i.e.,
$A_i^2=\eins$ for all $i$). Then $\sum_{i=1}^n \delta^2(A_i)\geq
n-1.$ To show this, is suffices to show that $\sum_{i=1}^n
\mean{A_i}^2 \leq 1.$ This can be proved as follows: We take real
coefficients $\lambda_1,...,\lambda_n$ with $\sum_{i=1}^n
\lambda_i^2=1.$ Then, using the fact that the $A_i$ anti-commute,
we have $(\sum_{i=1}^n \lambda_i A_i)^2 = \sum_{i=1}^n \lambda_i^2
A_i^2 = \sum_{i=1}^n \lambda_i^2\eins=\eins.$ So, for all states
$|\mean{\sum_{i=1}^n \lambda_i A_i}|=|\sum_{i=1}^n \lambda_i
\mean{A_i}| \leq 1$ holds. Since the $\lambda_i$ are arbitrary,
this implies that the vector of the mean values
$(\mean{A_1},\mean{A_2}, ... ,\mean{A_n})$ has a length smaller
than $1.$ Thus, $\sum_{i=1}^n \mean{A_i}^2 \leq 1$ has to hold.
This method can be used to derive all the bounds in Sec. IV.

%%%%%%%%%%%%%%%%%%%%%%%%%%%%%%%%%%%%%%%%%%%%%%%%%%%%%%%%%%%%%%%%%%%%%%


\begin{thebibliography}{99}
\bibitem{M90B} The Mermin inequality presented in this paper
is maximally violated by the state
$(\ket{0000...}+\ket{1111...})/\sqrt{2}.$ The form presented by
Mermin in Ref. \cite{M90} is maximally violated by
$(\ket{0000...}+i\ket{1111...})/\sqrt{2}.$ The two Bell
inequalities are equivalent under simple relabeling of variables.
\end{thebibliography}
\end{document}